\documentstyle[times,pramana,epsf,floats]{ias}
\begin{document}
\title{Crisis and unstable dimension variability in the bailout embedding map}

\author{N.Nirmal Thyagu and Neelima Gupte}
\address{Department of Physics, Indian Institute of Technology Madras, Chennai-600036}
\abstract{The dynamics of inertial particles in  $2-d$  incompressible flows can
be modeled by  $4-d$
bailout embedding maps. The density of the inertial particles, relative
to the density of the fluid, is a crucial parameter which controls
the dynamical behaviour of the particles.
We study here the dynamical behaviour of
aerosols, i.e. particles heavier
than the flow.
An attractor widening and merging
crisis is seen the phase space in the aerosol case.
Crisis induced intermittency is seen in the time series and
the laminar length distribution of times before bursts
gives rise to a power law with the exponent $\beta=-1/3$.
The maximum Lyapunov exponent near the crisis fluctuates
around zero indicating
unstable dimension variability (UDV) in the system.
The presence of unstable dimension variability
is confirmed by the behaviour of the probability distributions of
the finite time Lyapunov exponents.}

\maketitle
\section{Introduction}
The study of the dynamics of inertial particles in fluid flows is important  
from the  point of view of fundamental science, as well as that of engineering applications.
The  complex dynamics of aerosols in the atmosphere, plankton in oceans,
and impurity transport in industrial applications are a few of the varied contexts in which 
impurity dynamics is of interest.

The fluid flow, which is the base flow,  can be  modeled by a two-dimensional incompressible flow and
the Lagrangian dynamics of the particles, i.e. the impurities,  is given  by the Maxey-Riley
equation\cite{mot03,reig01,babia00}. This equation is further simplified to a minimal equation of motion
called the bailout embedding equation where the fluid flow dynamics is
embedded in a larger set of equations which includes the differences
between the particle and fluid velocities. Although the Lagrangian 
dynamics of the underlying fluid flow is incompressible, the particle
motion is compressible.     
A map analog of this equation, called the
embedding map \cite{mot03}, is constructed with base map given by an area preserving map. 
The dynamical behaviour of the inertial particles is determined by 
two parameters  $\alpha$ and $\gamma$, where $\alpha$ is the ratio of
densities of the particle and the fluid, and $\gamma$ is the dissipation
parameter. 
Here, the standard map is taken 
as the base map, which is a prototypical area preserving system. 

In this paper we confine our study to 
the aerosol regime of the embedding map i.e. the regime where the
particle density is greater than the fluid density. 
A crisis is seen in the phase space, and crisis induced intermittency is 
seen in the time series at certain $\alpha$ and $\gamma$ values. 
The time between subsequent bursts 
follows a scaling law as a function of the dissipation parameter, with the
exponent $\beta = -1/3$.
In the vicinity of the crisis, in the precrisis region, the maximum Lyapunov exponent  fluctuates
around zero. This is the characteristic indication of the presence
of unstable dimension variability (UDV) in the system. This is confirmed by
the   distribution 
of the finite time Lyapunov exponents (FTLE-s) obtained
in the precrisis and the post crisis scenarios. The FTLE-s in the
precrisis regime are distributed equally about zero confirming the
presence of UDV in the system in this
regime, whereas they 
shift to the positive side in the postcrisis regime.

\section{The Embedding Map}

The embedding equation which describes the inertial particle dynamics 
in a fluid flow is given by ,

\begin{eqnarray}
\frac{d{\bf v}}{dt} - \alpha \frac{d{\bf u}}{dt} & = & \gamma({\bf v - u}).
\end{eqnarray}

The velocity of the fluid parcel is given by ${\bf u}(x,y,t)$ and the particle
velocity is given by ${\bf v}$. The mass ratio parameter is
given by $\alpha = 3\rho_{f}/(\rho_{f}+2 \rho_{p})$, where
$\rho_{f}$ is the fluid density and $\rho_{p}$ is the 
particle density. The particles that are heavier than the fluid, i.e. the aerosols, 
correspond to  $\alpha < 1$ and the particles lighter than the fluid,
the bubbles,
correspond to  $\alpha >1$. The dissipation parameter $\gamma = 2\alpha/3St$, measures 
the expansion or contraction in the phase space, where $St$ is the 
Stokes number.

The map analog of the embedding equation is given by \cite{mot03},
\begin{eqnarray}
{\bf x}_{n+2} - {\bf M(x}_{n+1}) & = & e^{-\gamma}[\alpha{\bf x}_{n+1} - {\bf M(x}_{n})].
\end{eqnarray}
The base map defining the fluid flow is represented by an area preserving map ${\bf M}$ with 
the evolution given by, ${\bf x}_{n+1} =  {\bf M}({\bf x}_{n})$. As
mentioned above, while the base fluid flow
is incompressible, the advected particle motion is compressible and is therefore
represented by the dissipative embedding map.

We can rewrite the above equation as,
\begin{eqnarray}
{\bf x}_{n+1} &  = & {\bf M(x}_n)+{\bf \delta}_n \nonumber \\
{\bf \delta}_{n+1} & = & e^{-\gamma}[\alpha {\bf x}_{n+1}-{\bf M(x}_n)].
\label{bailtwo}
\end{eqnarray}

where a new variable ${\bf \delta}$ defines the detachment of the particle 
from the fluid. Here, the fluid dynamics is embedded in a larger phase space 
that 
describes the particle dynamics, and in the appropriate limit
$\gamma \to \infty$ and $\alpha = 1$, the detachment $\delta \to 0$ and 
the fluid dynamics is recovered. When $\delta$ is nonzero, the particle
detaches or bails out of the fluid trajectory. Therefore, this map
is called the bailout embedding map.

The standard map is taken as the base map ${\bf M(x}_{n})$, as it is a
widely used as a test bed for the study of mixing and transport of impurities
in fluids. The standard map is given by,

\begin{eqnarray}
{\bf x}_{n+1} & = & {\bf x}_n + {\bf y}_{n+1} \ ~~~~~~~~~~~ (Mod \  1)\nonumber \\
{\bf y}_{n+1} & = & {\bf y}_n +  \frac{K}{2\pi}\sin( 2\pi{\bf x}_n) \ ~~(Mod \  1)
\label{standard}
\end{eqnarray}

The chaoticity of the map is controlled by the nonlinearity parameter $K$.
Our study here will be confined to the  $K=2.0$ case, where we can see regions
of periodic islands, invariant curves and a  chaotic sea. 
The four dimensional standard embedding map ${\bf
M}^{'}(x_{n},y_{n},\delta_{x}(n),\delta_{y}(n))$ \cite{nir07}
is obtained by substituting the standard map for 
${\bf M(x_{n})}$ in the eqn.\ref{bailtwo}.

\begin{figure}[t]
\epsfxsize=55mm
\epsfysize=60mm
\begin{tabular}{cccc}
(a)&
\epsfbox{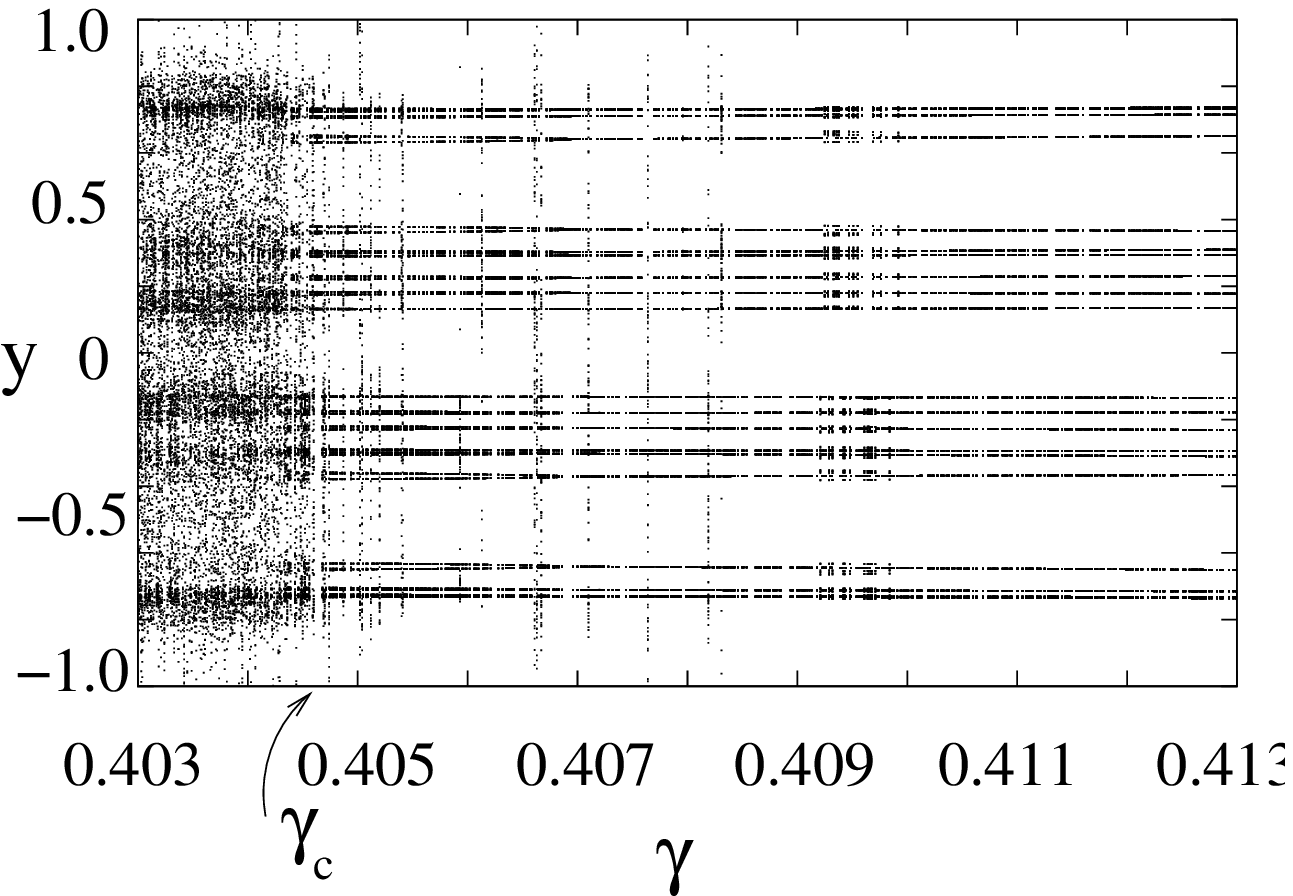}&
(b)&
\epsfxsize=55mm
\epsfysize=60mm
\vspace{5mm}
\epsfbox{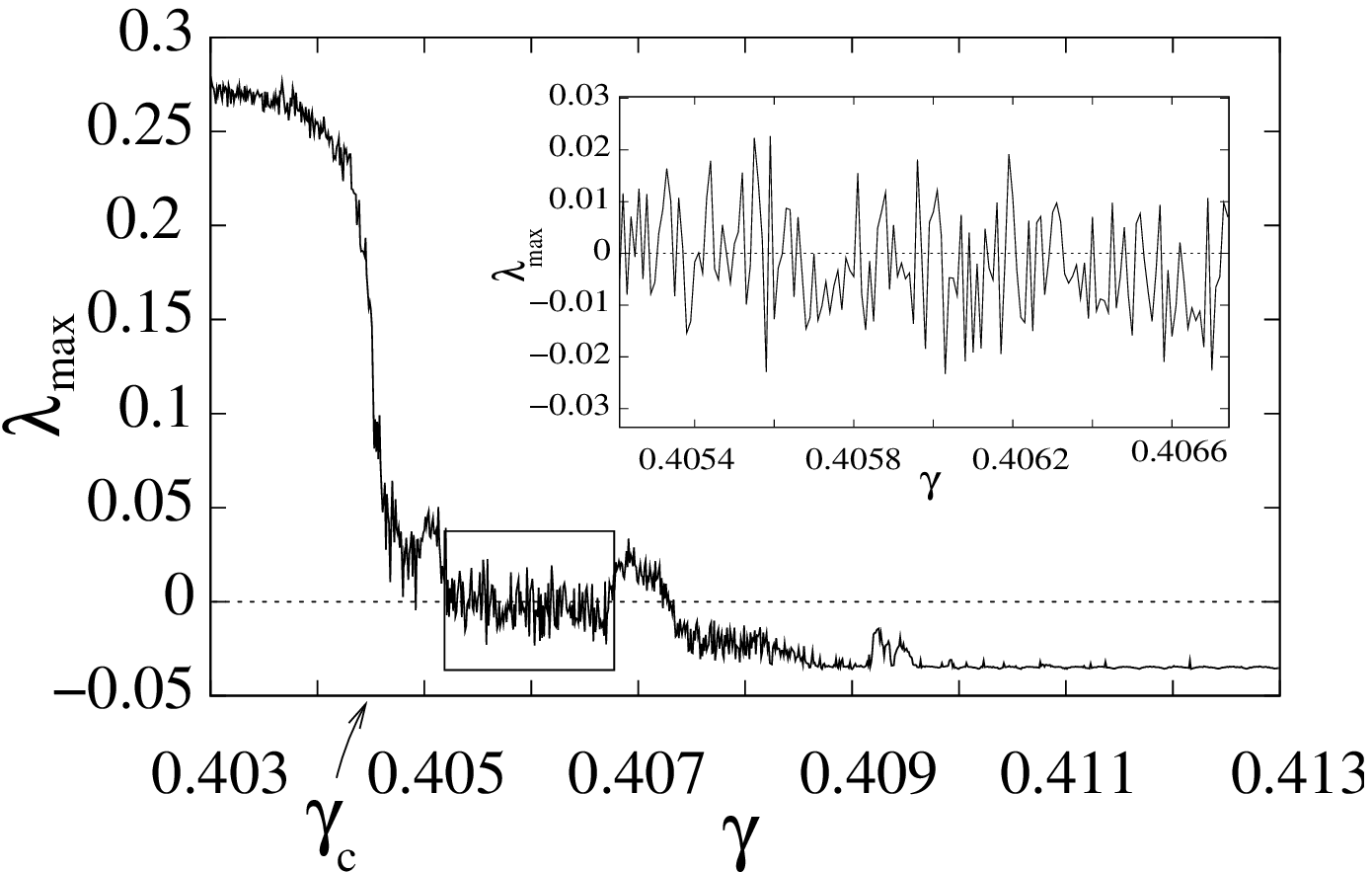}\\
\end{tabular}
\caption{At the critical dissipation parameter value $\gamma_{c}=0.40452$, 
(a) the bifurcation diagram shows attractor widening cum merging crisis and
(b) the maximum Lyapunov exponent plot shows a sudden jump. The  inset shows
the fluctuation of the $\lambda_{max}$ around zero in 
the range $\gamma=0.4052$ to $\gamma=0.4068$, where UDV is seen. Here
the transient is $2000$ iterates.}
\label{bif-lyap}
\end{figure}

\section{The crisis and intermittency in the aerosol regime}

The bifurcation diagram
of the embedding map in the aerosol region at $\alpha=0.8$ is plotted 
in Fig. 1(a). 
This  shows a sudden 
 change in the size of the attractor as the dissipation 
parameter is varied. Such a sudden change in the size
of the attractor is a signature of a crisis in the system.
From Fig. \ref{bif-lyap} (a) we see that as the dissipation
parameter $\gamma$ is decreased,  orbits of finite period 
widen suddenly into a large chaotic attractor at
$\gamma=\gamma_c=0.4045$.
 
\begin{figure}[t]
\epsfxsize=55mm
\epsfysize=60mm
\begin{tabular}{cccc}
(a)&
\epsfbox{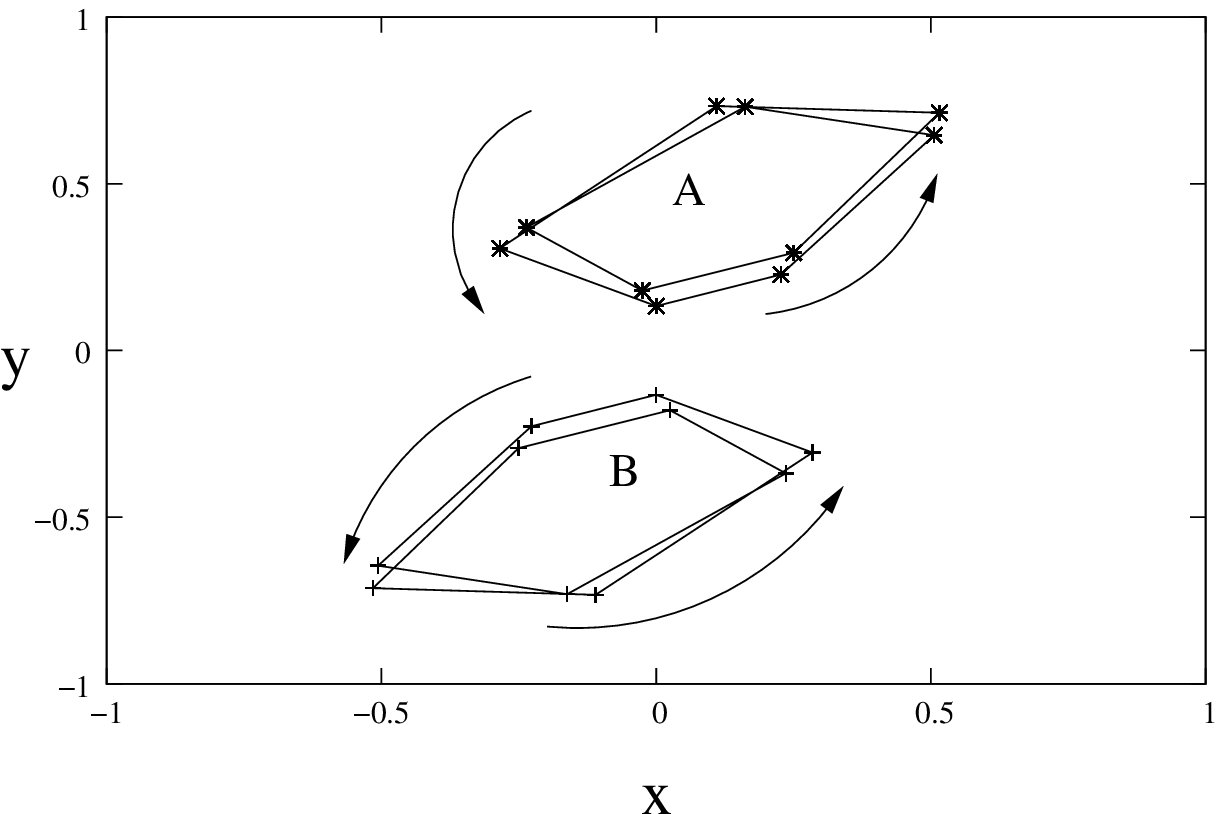}&

(b)&
\epsfxsize=55mm
\epsfysize=60mm
\epsfbox{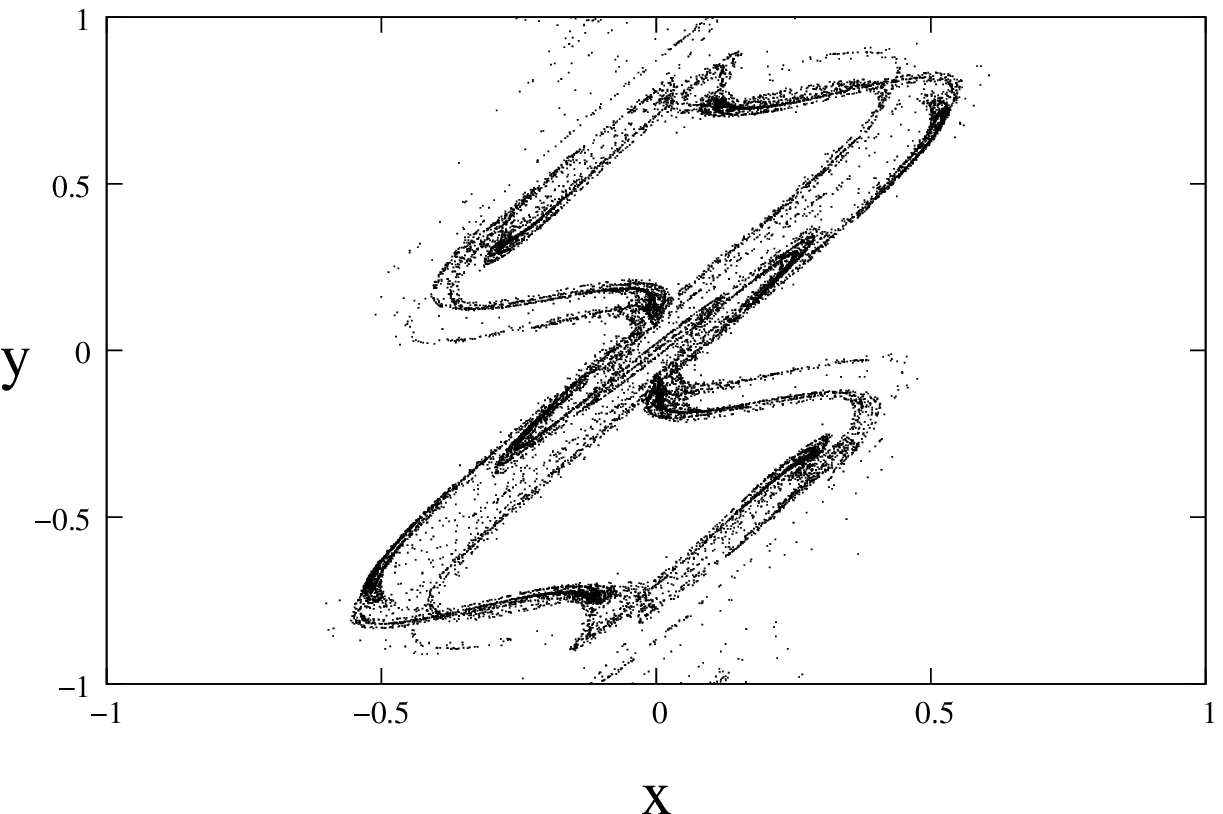}\\
\vspace{5mm}
\end{tabular}
\caption{The phase space plots of (a) the precrisis attractor ($\gamma=0.41$), (b)
the postcrisis attractor ($\gamma=0.40$).}
\label{phsp-cri}
\end{figure}

When the dissipation parameter is reduced from $\gamma=0.41$
through $\gamma=0.40$, two period-10 orbits in the phase space
merge and widen into a larger chaotic attractor (see Fig. \ref{phsp-cri}).
Therefore, the attractor merges and widens at the crisis.
The  value of $\gamma$ at which the crisis
occurs, viz. $\gamma=\gamma_c$,  can be identified from the bifurcation diagram. 
The maximum Lyapunov exponent, plotted in Fig. 
\ref{bif-lyap}(b), can also be
used as a quantifier for
finding the exact value of $\gamma_c$. At the crisis, i.e. at  the critical
parameter value  $\gamma_{c}$, the maximum Lyapunov exponent jumps suddenly.
The Lyapunov exponent is averaged over 100 initial conditions evolved
for  1000 time steps after the initial transient\cite{ic}.   
At the crisis, at the critical
parameter value  $\gamma_{c}=0.40452$ (see Ref. \cite{nir07} for a
plot to this accuracy in $\gamma$), $\lambda_{max}$ jumps suddenly as seen in Fig. \ref{bif-lyap}(b)).

\begin{figure}
\epsfxsize=50mm
\epsfysize=70mm
\begin{tabular}{cccc}
(a)&
\epsfbox{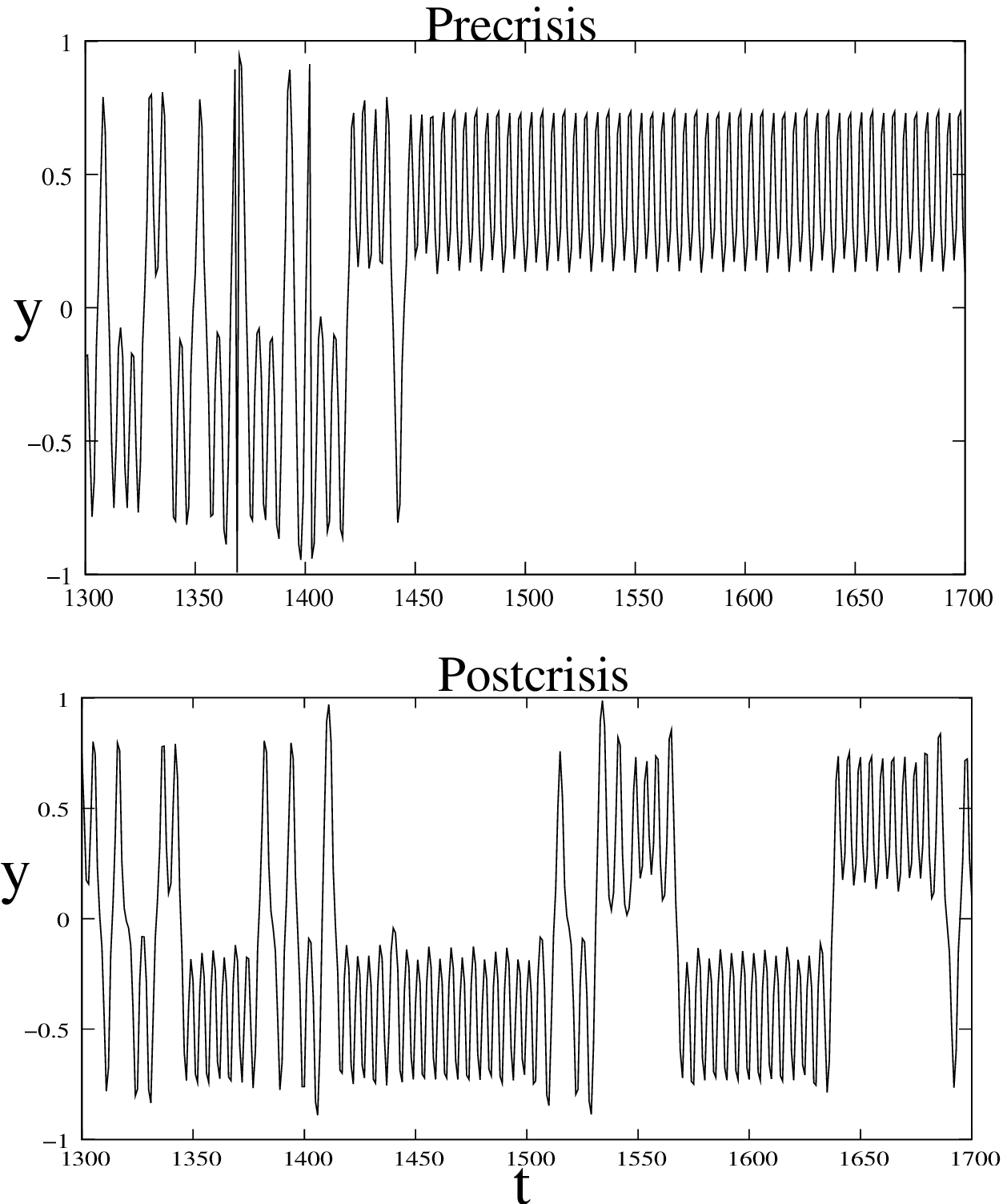}&
\hspace{5mm}
(b)&
\epsfxsize=50mm
\epsfysize=65mm

\epsfbox{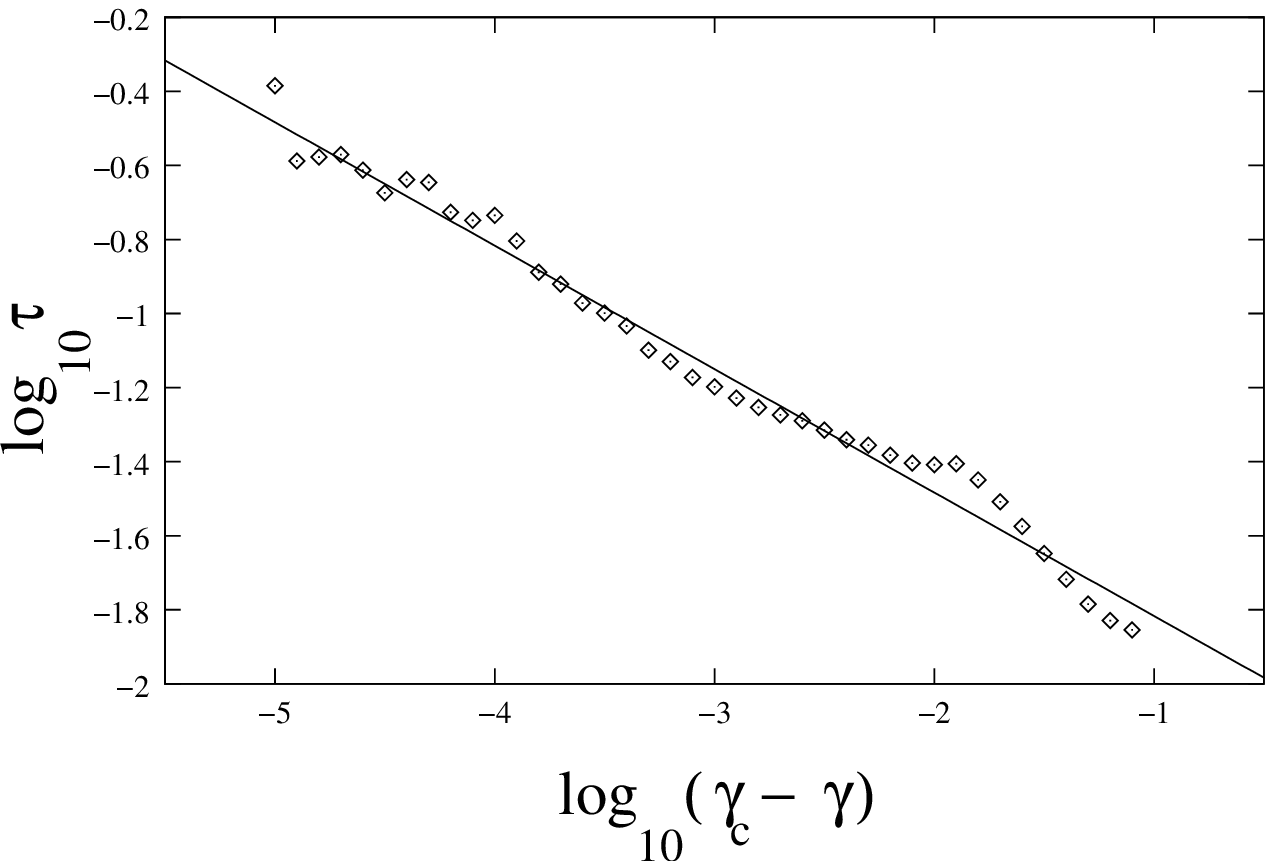}\\
\end{tabular}
\caption{(a) The time series of a typical trajectory in the precrisis ($\gamma=0.41$) and the postcrisis ($\gamma=0.40$).
(b) The log-log plot of the characteristic time $\tau$ versus $\gamma_{c} - \gamma$ with exponent $\beta = -1/3$.}
\label{scale}
\end{figure}

In the neighbourhood of the crisis we see crisis induced intermittency in the time series.
In Fig. \ref{scale} (a) the upper plot shows the time series in 
the precrisis situation. Here, the trajectory hops between the
two period-10 orbits and settles into 
one of them after the transient time. 
The lower plot in the same figure shows the time series in the post crisis situation.
Here, the trajectory hops between the two precrisis period-10 orbits
and the larger widened postcrisis attractor spending arbitrarily long
period of  time in each of them. 
To quantify the intermittency, a 
burst is said to have occurred when a trajectory staying in the precrisis  
orbits reaches  out to the postcrisis attractor. The time between 
the bursts is defined as the characteristic time $\tau$. In the vicinity
of the crisis ($\gamma_{c}$), 
 the average characteristic time $\tau$ is found to follow  power law
behaviour,
\begin{eqnarray}
\tau \sim (\gamma_{c}-\gamma)^\beta ,
\end{eqnarray}
where the exponent is found to be $\beta = -1/3 \pm 0.0527$. Fig \ref{scale}(b) shows 
the log-log plot of $\tau$ vs $(\gamma_{c} - \gamma)$.

\section{Presence of UDV} 
Near the critical dissipation parameter value $\gamma_{c}$, 
the maximum Lyapunov exponent plot shows that $\lambda_{max}$ 
fluctuates  near zero before it jumps suddenly to a positive
value.  The fluctuation of the maximum Lyapunov exponent 
between positive and negative values signals the presence
of unstable dimension variability in the system \cite{viana66_02}. This
phenomenon 
occurs in systems which have different numbers of  expanding
(unstable) directions. A typical trajectory evolving with time
passes through regions of the phase space having different 
number of expanding directions. For example, as the trajectory evolves, 
one or more  local Lyapunov exponents change signs
when going from  the regions having 
one unstable directions to the regions with more than
one unstable direction. The long time average of the 
Lyapunov exponents corresponding to the changing eigendirections
of a typical trajectory  will therefore average out to
have values near zero. 
The finite time Lyapunov exponents
(FTLE-s) can also be obtained for the above, and the distribution
of the  FTLE-s will have an equal share of both 
positive and negative regions corresponding to the 
expanding and contracting directions respectively, when
the UDV has maximum intensity in the system \cite{pereira17_07,yclai60_99}. 

The phenomenon of UDV has been discussed in a more general 
context where
the validity of the computer generated  numerically
trajectories,  also called psuedotrajectories,
is called into question \cite{ydo69_04}. More precisely,
when a system shows UDV, it belongs to a class of 
nonhyperbolic dynamical systems where the 
pseudotrajectories may not
shadow the true trajectories of the system for long times.
In contrast, the pseudotrajectories of the hyperbolic
systems will shadow the  true trajectories
for long times. 
Nonhyperbolicity in low dimensional systems appears
as tangencies between the stable and unstable manifolds. 
In higher dimensional systems, nonhyperbolicity manifests
itself as unstable dimension variability. 

\begin{figure}
\epsfxsize=80mm
\epsfysize=60mm
\begin{center}
\epsfbox{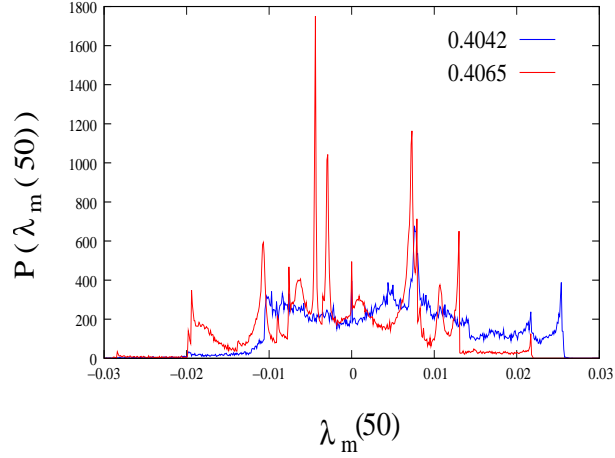}\\
\end{center}
\caption{(Color Online)The probability distribution of the finite time Lyapunov
exponents for $t=50$ for the pre-crisis ($\gamma=0.4065$, red colour)
region and the 
postcrisis  ($\gamma=0.4042$, blue colour) parameter values. The FTLE-s
are equally distributed about zero in the precrisis case indicating the presence of
UDV in the region.}
\label{ftle}
\end{figure}

In order to confirm the presence of UDV in our system, we
further obtain the finite time Lyapunov exponents
in the precrisis and postcrisis regions. 
The time-n finite time Lyapunov exponent in the eigendirection
$x$ is defined as,

\begin{eqnarray}
\lambda_{x} (x_{0},y_{0},\delta_{x}(0),\delta_{y}(0)) = \frac{1}{n} \displaystyle\sum_{i=1}^n ln \left | \frac{\partial {\bf M}^{'}(x_{n},y_{n},\delta_{x}(n),\delta_{y}(n))}{\partial x_{i}} \right | ,
\end{eqnarray}	

where ${\bf M}^{'}$ is the embedding map defined above. Here, $n=50$. 
In Fig. \ref{ftle},  $\gamma=0.4065$ (red colour) case
corresponds to the precrisis situation, and the distribution
has regions spread equally in positive and negative regions. 
Thus the distribution of FTLE-s confirms the presence of 
unstable dimension variability in the precrisis region.
This is also consistent with the observation that the largest Lyapunov
exponent fluctuates about zero in Fig. \ref{bif-lyap} (b) in the parameter
regime  $\gamma=0.4052$ to $\gamma=0.4068$. 
In contrast,  the $\gamma=0.4042$ (blue colour) case in Fig. \ref{ftle}
which shows the FTLE distribution in the postcrisis
situation (i.e. below $\gamma=0.40452$, has  shifted to the positive
region. As expected, the corresponding 
long time average  Lyapunov exponent at $\gamma=0.4042$ 
takes a positive value (see Fig. \ref{bif-lyap}(b)) as well.

As mentioned earlier, the presence of UDV in the embedding map for 
certain parameter ranges of dissipation, seen here for the aerosol case, 
can imply the breakdown of  shadowing. This can have
important implications for the behaviour of inertial particles. 
We plan to explore these in further work.

\section{Conclusion}
The embedded  standard map  is used to model the inertial particle dynamics
in a $2-d$ incompressible flow. We concentrate on
the dynamical behaviour in the aerosol region. A crisis is seen in 
in the bifurcation diagram and  the largest Lyapunov exponent plot
identifies 
the precise value of $\gamma$ at which the crisis occurs.
The statistics of characteristic times between the bursts  
corresponding to the crisis induced intermittency,
behaves as a power law with the exponent $\beta=-1/3$.

The fluctuation of the largest Lyapunov exponent $\lambda_{max}$ around zero, in the neighbourhood of
the  
crisis in the precrisis region, indicates  the possibility of the existence  
of unstable dimension variability in the system. Further, the finite time 
Lyapunov exponent distribution in the  precrisis and the postcrisis
scenarios, confirms the presence of UDV in the precrisis region. 
Thus the shadowing times for the embedded map are expected to be short
near the crisis. The implications of this for the dynamics of inertial 
particles deserve to be explored further. The consequences for aerosols,
as in the present study, can be of interest in realistic application
contexts. We hope to study these consequences, as well as the role of
unstable periodic orbits in the UDV in future work.


\begin{thebibliography}{99}
\bibitem{ic} Similar results are seen for single initial conditions,
including the single initial condition  used for plotting the bifurcation diagram.
\end{thebibliography}
\end{document}